\documentclass[12pt,a4paper]{article}
\usepackage{amssymb}

\usepackage[left=5em]{geometry}
\usepackage{amsmath}
\usepackage{graphicx}

\title{\textbf{Weak Gravity Conjecture and Holographic/Agegraphic Dark Energy}}
\author{Cheng-Yi Sun\footnote{cysun@mailis.gucas.ac.cn; ddscy@163.com}\
$^{,1}$\ and Rui-Hong Yue$^{2}$
\\
 {$^1$\small Institute of Modern Physics, Northwest University, Xian 710069, P.R.
 China.}\\
{$^2$\small Faculty of Science, Ningbo University, Ningbo 315211,
P.R. China.}}

\begin{document}
\maketitle
\begin{abstract}
A criterion that should be satisfied in the inflation and
quintessence models has been motivated from the weak gravity
conjecture (WGC) by Huang . However, it is found that the criterion
is inconsistent with the Holographic dark energy (HDE) and new
agegraphic dark energy (NADE) models. In the note, we firstly show
that in the HDE and NADE models the criterion should be be replaced
respectively by two new criterions. Secondly, we apply the new
criterions indicated by WGC to survey the two models. We find that
the contradiction between WGC and the NADE model is removed when the
new criterion is used. In the HDE model, we find the effects of the
spatial curvature and the interaction should be considered in order
to match the new criterion.
\end{abstract}



\section{Introduction}
A full quantum theory of gravity has not been known, but it is
generally believed that the theory of quantum gravity can be
formulated in the context of string theory. Recent progress
\cite{kklt} suggests that there exist a vast number of
semi-classical consistent vacua in string theory, named
\emph{Landscape} \cite{landscape}. However, not all semi-classical
consistent vacua are actually consistent on the quantum level, and
those actually inconsistent vacua are called \emph{Swampland}
\cite{vafa}. Self-consistent landscape is surrounded by the
swampland. Then, how to select the consistent landscape from the
numerous vacua becomes an urgent problem.

Recently,  The weak gravity conjecture (WGC) is suggested to be a
new criterion to distinguish the landscape from the swampland
\cite{wgc1,wgc2}. The conjecture can be most simply stated as
gravity is the weakest force. For a four-dimensional U(1) gauge
theory, WGC implies that there is an intrinsic UV cutoff \cite{wgc1}
\[
  \Lambda\leq gM_p,
\]
where $g$ is the gauge coupling constant and $M_p$ is the Planck
scale. In \cite{0703071}, the conjecture is applied to the scalar
field theory and it is argued that WGC also indicates an intrinsic
UV cutoff for the scalar field theories with gravity, e.g.
\[
  \Lambda\leq \lambda^{1/2} M_p
\]
for $\lambda\phi^4$ theory. In the slow-rolling inflation model with
the potential $V(\phi)\sim\lambda\phi^4$, it is natural to take the
Hubble parameter $H$ as the IR cutoff for the field theory. Then the
requirement that the IR cutoff should be lower than the UV cutoff
indicates \cite{0703071}
\begin{equation}
  \label{HLmabda}\frac{\lambda^{1/2}\phi^2}{M_p}\sim H\leq\Lambda\leq \lambda^{1/2}
  M_p, \quad\texttt{or,}\quad  \phi\leq M_p
\end{equation}
This leads Huang in \cite{0706.2215} to conjecture that the
variation of the inflaton during the period of inflation should be
less than $M_p$,
\begin{equation}
  \label{deltaPhiMp}|\Delta\phi|\leq M_p.
\end{equation}
And it is found that this can make stringent constraints on the
spectral index of the inflation model \cite{0706.2215}.

Furthermore, by arguing that the variation of the canonical
quintessence field minimally coupled to gravity should also be less
than the Planck scale, the author in \cite{0708.2760} used the
criterion (\ref{deltaPhiMp}) to explore the quintessence model of
dark energy, and found that the theoretic constraints are more
stringent than present experiments in some cases \cite{0708.2760}.

Recently, the criterion (\ref{deltaPhiMp}) has been used to explore
the different models of dark energy
\cite{0710.1406,0709.1517,0806.2415,1007.1517,1005.2466,0811.1333}.
By assuming that the holographic dark energy (HDE) scenario
\cite{HDE} is the underlying theory of dark energy and can be
described by the low-energy scalar field \cite{HQDE}, the criterion
is used in \cite{0709.1517,0806.2415,1007.1517} to survey the
holographic dark energy model. And similarly, the criterion
(\ref{deltaPhiMp}) is also used in \cite{1005.2466} to survey the
new agegraphic dark energy (NADE) \cite{0708.0884}. However, it is
found in \cite{0806.2415} that the criterion (\ref{deltaPhiMp})
cannot be satisfied in the non-interacting HDE mode without spatial
curvature. For the interacting HDE model with spatial curvature, it
is shown in \cite{1007.1517} that it is only on the edge of the
parameter space that is possible for WGC to be satisfied. For the
NADE model, it is found in \cite{1005.2466} that the criterion
(\ref{deltaPhiMp}) cannot be satisfied, either.

The tension between WGC and the two important dark energy models
makes us nervous. Motivated by this, in the note, we will try to
alleviate the tension by suggesting two new criterions indicated by
WGC to replace Eq.(\ref{deltaPhiMp}) in the HDE and NADE model
respectively.

The note is organized as follows. In the next section, we will first
give our motivation, and then suggest two new criterions to replace
Eq.(\ref{deltaPhiMp}) in the HDE and NADE models respectively. In
Sec.\ref{SecADE}, we will use the new criterion to survey the ANDE
model. In Sec.\ref{SecHDE}, we will use the new criterion to survey
the HDE model. In Sec.\ref{SecCD}, Conclusion and Discussion will be
given.

\section{New Criterions of WGC in HDE and NADE Models}
\label{SecNewCriterion}

In the last section, we have recalled how the criterion
(\ref{deltaPhiMp}) is motivated in \cite{0703071,0706.2215}. A key
assumption in the motivation is that the IR cutoff scale is taken to
be the Hubble parameter, $H$. In the inflation models and the
quintessence models, the assumption is reasonable. However, for the
HDE and NADE models, the assumption contradicts the basic motivation
of the two models. This can be understood easily. We know that the
HDE model arises from the holographic principle \cite{HP} which
imposes a relation between the UV cutoff and IR cutoff on the
effective quantum field theory. Then, motivated by the principle,
the energy density of HDE, $\rho_d$, is taken to be
\begin{equation}
  \label{HDE}
  \rho_d=\frac{3c^2M_p^2}{L^2},
\end{equation}
where $M_p=(8\pi G)^{-1/2}$, $c$ is a constant, and $L$ is the IR
cutoff length scale which is proposed to be the future event horizon
$R_h$ \cite{HDE}
\begin{equation}
  \label{FEH}
  R_h=a(t)\int^{+\infty}_t{\frac{dt'}{a(t')}}.
\end{equation}
Then, clearly, in the HDE model, the IR cutoff is not the Hubble
scale, but the future event horizon. Actually, it has been pointed
out in \cite{0403052} that the model with $L=H^{-1}$ does not work.

The NADE model can be taken as a different version of the HDE model
by choosing $L$ to be the conformal age of the universe. The energy
density of NADE, $\rho_q$, is proposed to be \cite{0708.0884}
\begin{equation}
  \label{NADE}
  \rho_q=\frac{3n^2M_p^2}{L^2},
\end{equation}
where $n$ is a constant, and the IR cutoff length scale $L$ is taken
to be the conformal age of the universe $\eta$ \cite{0708.0884}
\begin{equation}
  \label{confAge}
  \eta=\int^t_0{\frac{dt'}{a(t')}}.
\end{equation}
Then in the NADE model, the IR cutoff is also not the Hubble scale,
but the conformal age.

Since the Hubble scale can not be taken to be the IR cutoff scale in
the HDE and NADE models, we conclude that the criterion
(\ref{deltaPhiMp}) cannot be applied to constrain the two models.
Then, in order to explore the constraints imposed by WGC on the HDE
and NADE models, we need to find a new criterion. To do so, we still
follow the analysis of Huang \cite{0703071,0706.2215}. Considering a
slow-rolling quintessence field with the the potential $V(\phi)\sim
\lambda\phi^4$, we get an intrinsic UV cutoff indicated by WGC
\cite{0703071}
\begin{equation}
  \label{Lambda}
  \Lambda\le\lambda^{-1/2}M_p.
\end{equation}
Then, by assuming that HDE can be described effectively by the
quintessence field, we have
\begin{equation}
  \label{LC}
  \rho_d\simeq V(\phi)\sim \lambda\phi^4\Rightarrow L^{-2}\sim
  \frac{\lambda\phi^4}{c^2M_p^2}.
\end{equation}
Here the IR cutoff length scale for the scalar field theory should
be the future event horizon $L$ as given in Eq.(\ref{FEH}). Then the
requirement that the IR cutoff should be lower than the UV cutoff
tells us that
\begin{equation}
  \label{LLambda}
  L^{-1}\le\Lambda\le\lambda^{-1/2}M_p.
\end{equation}
Together with Eq.(\ref{LC}), we get
\begin{equation}
  \label{phiWGC}
  |\phi|\le cM_p.
\end{equation}
Then from the result, we may naturally conjecture a new criterion to
replace Eq.(\ref{deltaPhiMp}) in the HDE model that the variation of
the holographic quintessence field should be less than $cM_p$,
\begin{equation}
  \label{newDeltaPhiHDE}
  |\Delta\phi|\le cM_p.
\end{equation}
Since $c\ge1$ in the holographic quintessence model, the new
criterion (\ref{newDeltaPhiHDE}) is easier to be satisfied than
Eq.(\ref{deltaPhiMp}).

Similarly, for the NADE model, a new criterion can also be motivated
that the variation of the agegraphic quintessence field should be
less than $nM_p$,
\begin{equation}
  \label{newDeltaPhiNADE}
  |\Delta\phi|\le nM_p.
\end{equation}
Obviously, the new criterion (\ref{newDeltaPhiNADE}) is much looser
than Eq.(\ref{deltaPhiMp}), since the analysis of the observational
data tells us that $n>2.6$ \cite{0904.0928,1011.6122}. Below we will
use the criterions (\ref{newDeltaPhiHDE}) and
(\ref{newDeltaPhiNADE}) to survey the HDE and NADE models
respectively.

\section{Agegraphic Quintessence Model and New Criterion of WGC}
\label{SecADE}

In the section, we will use the criterion (\ref{newDeltaPhiNADE}) to
survey the NADE model. The fractional energy density of NADE is
given by
\begin{equation}
  \label{FracNADE}
  \Omega_q=\frac{n^2}{H^2\eta^2},
\end{equation}
where $H=\dot{a}/a$, $a(t)$ is the scale factor in the
Friedmann-Robertson-Walker (FRW) metric and a dot denotes the
derivative with respect to the cosmic time $t$. Then using
Eqs.(\ref{NADE}), (\ref{confAge}), (\ref{FracNADE}) and the
conservation law of NADE
\begin{equation}
  \label{CLNADE}
  \dot{\rho}_q+3H(1+w_q)\rho_q=0,
\end{equation}
we obtain the equation of state parameter of NADE \cite{0708.0884}
\begin{equation}
  \label{EOSNADE}
  w_q=-1+\frac{2}{3n}\frac{\sqrt{\Omega_q}}{a}.
\end{equation}
For a flat FRW universe filled by the NADE and the pressureless
matter, the Friedmann equation reads
\begin{equation}
  \label{FredEqNADE}
  H^2=\frac{1}{3M_p^2}(\rho_q+\rho_m),
\end{equation}
where $\rho_m$ is the energy density of matter with the conservation
law
\begin{equation}
  \label{CLMatter}
  \dot{\rho}_m+3H\rho_m=0.
\end{equation}
Then, from Eq.(\ref{FracNADE}) and using Eqs.(\ref{CLNADE}),
(\ref{FredEqNADE}) and (\ref{CLMatter}), we can get the evolving
equation of $\Omega_q$ \cite{0708.0884}
\begin{equation}
  \label{EqOmegaNADE}
  \Omega'_q=-\Omega_q(1-\Omega_q)\Big(\frac{3}{1+z}-\frac{2}{n}\sqrt{\Omega_q}\Big),
\end{equation}
where $\Omega'_q\equiv\frac{d\Omega_q}{dz}$, and $z=\frac{a_0}{a}-1$
is the cosmological redshift. In the note, the subscript $0$ denotes
the present value of the corresponding parameter,  and we take
$a_0=1$.

Now considering a single-scalar-field quintessence model with the
potential $V(\phi)$, we assume that the field is spatially
homogeneous. Then the energy density and pressure of the
quintessence scalar field are
\begin{align}
  \label{phiED}
  \rho_\phi&=\frac{1}{2}\dot{\phi}^2+V(\phi),\\
  \label{phiP}
  p_\phi&=\frac{1}{2}\dot{\phi}^2-V(\phi).
\end{align}
Then we can obtain easily
\begin{equation}
  \label{rhoWphi}
  \rho_\phi=\frac{\dot{\phi}^2}{1+w_\phi},
\end{equation}
where $w_\phi=\frac{p_\phi}{\rho_\phi}$. Without loss of generality,
we may assume $dV/d\phi<0$ and $\dot{\phi}>0$. Thus from
Eq.(\ref{rhoWphi}), we may have
\begin{equation}
  \label{dPhidt}
  \dot{\phi}=\sqrt{(1+w_\phi)\rho_\phi}
\end{equation}
Then the criterion (\ref{newDeltaPhiHDE}) tells us
\begin{equation}
  \label{QuinWGCHDE}
  c\geq\frac{|\Delta\phi(z)|}{M_p}=\int{\frac{\dot{\phi}}{M_p}dt}=\int^z_0{\sqrt{3[1+w_\phi(z')]\Omega_\phi(z')}\frac{dz'}{1+z'}},
\end{equation}
where $\Omega_\phi=\rho_\phi/(3H^2M_p^2)$.  And the criterion
(\ref{newDeltaPhiNADE}) tells us
\begin{equation}
  \label{QuinWGCNADE}
  \begin{split}
    n\geq\frac{|\Delta\phi(z)|}{M_p}=\int{\frac{\dot{\phi}}{M_p}dt}=\int^z_0{\sqrt{3[1+w_\phi(z')]\Omega_\phi(z')}\frac{dz'}{1+z'}},
  \end{split}
\end{equation}

\begin{figure}
\centering
\renewcommand{\figurename}{Fig.}
\includegraphics[scale=0.6]{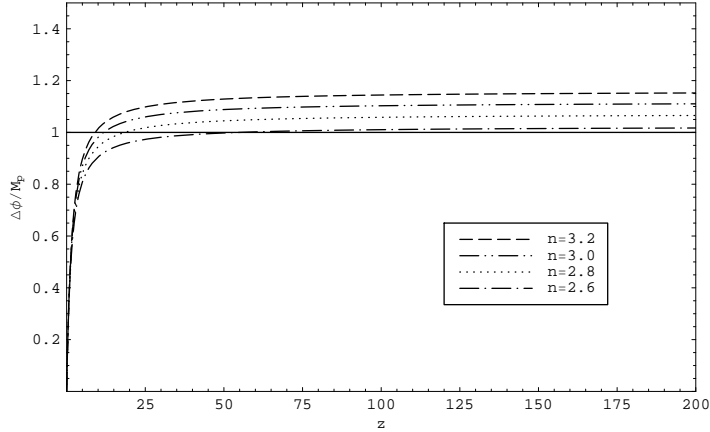}
\caption{$\Delta\phi(z)/M_p$ versus the redshift $z$ in the
agegraphic quintessence model with the initial condition
 $\Omega_q(z_i)=\frac{n^2}{4(1+z_{i})^2}$ at $z_i=2000$. Obviously, $n\ge\Delta\phi(z)/M_p$ holds.\label{FigNADE}}
\end{figure}

By assuming NADE can be describe effectively by the quintessence
field, we can have
\begin{equation}
  \label{QuinNADE}
  \rho_\phi=\rho_q\Rightarrow\Omega_\phi=\Omega_q,\quad w_\phi=w_q.
\end{equation}
By substituting the equations into Eq.(\ref{QuinWGCNADE}), finally
we obtain the constraint imposed by WGC on NADE
\begin{equation}
  \label{NADEWGC}
  n\geq\frac{|\Delta\phi(z)|}{M_p}=\int^z_0{\sqrt{\frac{2\Omega_q^{3/2}}{n(1+z)}}dz},
\end{equation}
where Eq.(\ref{EOSNADE}) has been used. Since $\Omega_q$ can be
obtained by solving Eq.(\ref{EqOmegaNADE}) numerically with the
initial condition $\Omega_q(z_i)=\frac{n^2}{4(1+z_{i})^2}$ at
$z_i=2000$ \cite{0708.0884}, it is easy for us to check whether NADE
is consistent with the condition (\ref{NADEWGC}). We display the
results in Fig.\ref{FigNADE}. It should be noted that the NADE model
has been constrained strictly by using the latest observational
data. By analyzing the observational data, it is shown in
\cite{0904.0928} that $n=2.807^{+0.087}_{-0.086}$ at the $68.3\%$
confidence level, and $n=2.807^{+0.176}_{-0.170}$ at the $95.4\%$
confidence level. And it is shown in \cite{1011.6122} that
$n=2.886^{+0.084}_{-0.082}$ at $1\sigma$ confidence level, and
$n=2.886^{+0.169}_{-0.163}$ at $2\sigma$ confidence level. Then from
Fig.\ref{FigNADE}, we know that, in the NADE model, the new
criterion (\ref{newDeltaPhiNADE}) is consistent with the current
observational constraints.

\section{Holographic Quintessence Model and New Criterion of WGC}
\label{SecHDE}

Since the HDE model has been analyzed in detail by using the
criterion (\ref{deltaPhiMp}) in \cite{1007.1517}, we can apply the
new criterion (\ref{newDeltaPhiHDE}) to constrain the HDE model just
by repeating the analysis in \cite{1007.1517} simply. Then in the
section, we will only display the results summarily.

Considering the FRW universe filled by HDE and pressureless matter,
the corresponding Friedmann equation reads
\begin{equation}
  \label{FredEqNADE}
  H^2+\frac{k}{a^2}=\frac{1}{3M_p^2}(\rho_d+\rho_m),
\end{equation}
where $k$ is the curvature parameter with $k=-1,0,1$ corresponding
to a spatially open, flat and closed universe respectively, and
$\rho_m$ is the energy density of matter. Usually, by defining
\begin{equation}
  \label{fractionalED}
  \Omega_k=\frac{\rho_k}{\rho_c}=\frac{k}{H^2a^2},\quad \Omega_d=\frac{\rho_d}{\rho_c},
  \quad \Omega_m=\frac{\rho_m}{\rho_c},
\end{equation}
where $\rho_k=k/a^2$ and $\rho_c=3M_p^2H^2$, we can rewrite the
Friedmann equation as
\begin{equation}
  \label{reFD}
  1+\Omega_k=\Omega_d+\Omega_m.
\end{equation}
In the universe with the spatial curvature, the future event horizon
$R_h$ should be defined as $R_h(t)=a(t)r(t)$ and $r(t)$ satisfying
\cite{0404229}
\begin{equation}
  \label{FEHSpatCurv}
  \int^{r(t)}_0{\frac{dr}{\sqrt{1-kr^2}}}=\int^\infty_t{\frac{dt'}{a(t')}}.
\end{equation}
By defining the effective equation of state parameter as
$\dot{\rho}_d+3H(1+w^{\text{eff}}_d)\rho_d=0$, we may get
\cite{1007.1517}
\begin{equation}
  \label{weff}
  w^{\text{eff}}_d=-\frac{1}{3}\Big(1+2\sqrt{\frac{\Omega_d}{c^2}-\Omega_k}\Big).
\end{equation}

The interaction between HDE and matter can be described by the
conservation laws
\begin{align}
  \label{CLMatterQ}
  \dot{\rho}_m+3H\rho_m&=Q,\\
  \label{CLHDEQ}
  \dot{\rho}_d+3H(1+w_d)\rho_d&=-Q,
\end{align}
where $Q$ denotes the phenomenological interaction term. We,
following Ref.\cite{0910.3855}, consider three types of interaction
\begin{align}
  \label{Q1}
  Q_1&=-3b\rho_d\\
  \label{Q2}
  Q_2&=-3b(\rho_d+\rho_m)\\
  \label{Q3}
  Q_3&=-3b\rho_m
\end{align}
For convenience, we uniformly express the interaction term as
\begin{equation}
  \label{Qi}
  Q_i=-3bH\rho_c\Omega_i,
\end{equation}
where $\Omega_i =\Omega_d,1+\Omega_k$ and $\Omega_m$, for $i=1,2$
and $3$, respectively.

Then, by using Eqs.(\ref{HDE}), (\ref{reFD}), (\ref{FEHSpatCurv}),
(\ref{CLMatterQ}), (\ref{CLHDEQ}) and (\ref{Qi}), we get the
evolving equations of the interacting HDE with spatial curvature
\cite{0910.3855}
\begin{align}
  \label{dHdz}
  \frac{d\widetilde{H}}{dz}&=-\frac{\widetilde{H}}{1+z}\Omega_d\Bigg(\frac{3\Omega_d-\frac{\Omega_{k0}(1+z)^2}{\widetilde{H}^2}-3-3b\Omega_i}{2\Omega_D}-1
                                                                    +\sqrt{\frac{\Omega_d}{c^2}-\frac{\Omega_{k0}(1+z)^2}{\widetilde{H}^2}}\Bigg),\\
  \label{dOmegaHDEdz}
  \frac{d\Omega_d}{dz}&=-\frac{2\Omega_d(1-\Omega_d)}{1+z}\Bigg(\sqrt{\frac{\Omega_d}{c^2}-\frac{\Omega_{k0}(1+z)^2}{\widetilde{H}^2}}-1
                                                             -\frac{3\Omega_d-\frac{\Omega_{k0}(1+z)^2}{\widetilde{H}^2}-3-3b\Omega_i}{2(1-\Omega_d)}\Bigg),
\end{align}
where $\widetilde{H}\equiv\frac{H}{H_0}$.

Similarly to the case of NADE, by assuming that HDE can be described
effectively by the quintessence field with potential $V(\phi)$, we
can have
\begin{equation}
  \label{QuinHDE}
  \rho_\phi=\rho_d\Rightarrow\Omega_\phi=\Omega_d,\quad w_\phi=w_d^{\text{eff}}.
\end{equation}
Then substituting the equations into Eq.(\ref{QuinWGCHDE}) and using
Eq.(\ref{weff}), we get the new constraint imposed by WGC on HDE as
\begin{equation}
  \label{HDEWGC}
  \begin{split}
    1\geq\frac{|\Delta\phi(z)|}{cM_p}=\int^z_0{\sqrt{2\Big(1-\sqrt{\frac{\Omega_d(z')}{c^2}
                                                           -\frac{\Omega_{k0}(1+z')^2}{\widetilde{H}^2}}\Big)\Omega_d(z')}\frac{dz'}{(1+z')c}}.
  \end{split}
\end{equation}
By solving Eqs.(\ref{dHdz}) and (\ref{dOmegaHDEdz}) numerically to
obtain $\Omega_d(z)$, we can easily check whether the constraint is
satisfied in the HDE scenario.

\begin{figure}
\centering
\renewcommand{\figurename}{Fig.}
\includegraphics[scale=0.6]{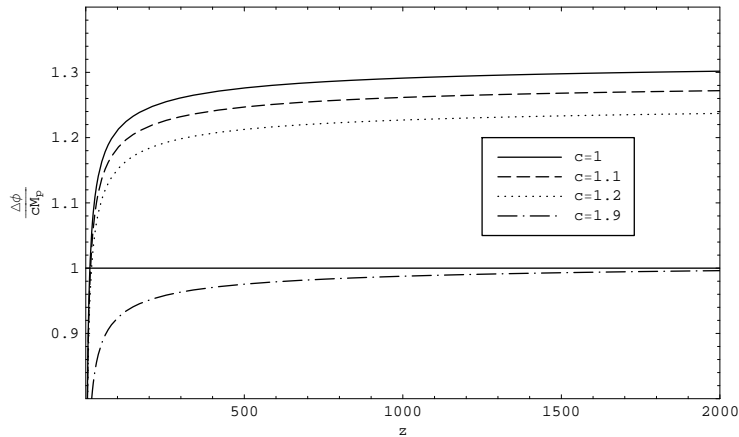}
\caption{$\frac{\Delta\phi(z)}{cM_p}$ versus the redshift $z$ in the
non-interacting holographic quintessence model in the flat universe
with the fixed initial condition $\Omega_{m0}=0.34$ for different
$c$. \label{FigHDE}}
\end{figure}

\begin{figure}
\centering
\renewcommand{\figurename}{Fig.}
\includegraphics[scale=0.6]{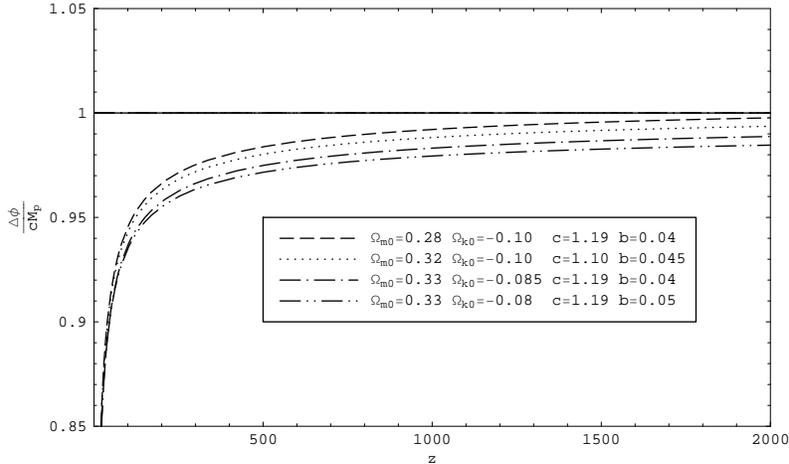}
\caption{$\frac{\Delta\phi(z)}{cM_p}$ versus the redshift $z$ in the
holographic quintessence model with spatial curvature and
interaction term $Q=Q_3$. \label{FigKIHDE3}}
\end{figure}

The result of the HDE model without interaction and spatial
curvature is displayed in Fig.\ref{FigHDE}. We find that the new
criterion is still inconsistent with the non-interacting HDE model
in the flat universe. Naively, in order to match WGC, we should take
$c\gtrsim 1.9$ which has been far outside the parameter range
obtained in \cite{0904.0928}. In Fig.\ref{FigHDE}, we have fixed
$\Omega_{m0}=0.34$, since smaller $\Omega_{m0}$ would make it more
difficult for Eq.(\ref{HDEWGC}) to be satisfied and the value has
been bigger than the the maximum value of $\Omega_{m0}$ in
observational range \cite{0904.0928}. Actually, we find that it is
only in the HDE model with spatial curvature and interaction $Q=Q_3$
that it is possible for Eq.(\ref{HDEWGC}) to be satisfied. Since the
conclusion is similar to that in \cite{1007.1517}, here we only
display the result of the HDE model with spatial curvature and
interaction $Q=Q_3$ in Fig.\ref{FigKIHDE3}. Roughly, we find that if
$b\gtrsim0.04$ and $\Omega_{k0}\lesssim-0.08$, it is possible to
find the combinations of $\Omega_{m0},\Omega_{k0},b$ and $c$ which
satisfy Eq.(\ref{HDEWGC}). The parameter range within which
Eq.(\ref{newDeltaPhiHDE}) can be satisfied is larger than that
obtained in \cite{1007.1517} with Eq.(\ref{deltaPhiMp}).

\section{Conclusion and Discussion}
\label{SecCD}

Recently, WGC is suggested to be a new criterion to distinguish the
landscape from the swampland \cite{wgc1,wgc2}. From the conjecture,
Huang \cite{0703071,0706.2215} motivates a constraint
(\ref{deltaPhiMp}) that should be satisfied in the inflation and
quintessence models. However, it is found that Eq.(\ref{deltaPhiMp})
can not be satisfied in the agegraphic quintessence model
\cite{1005.2466}. And it is very difficult for Eq.(\ref{deltaPhiMp})
to be satisfied in the holographic quintessence model
\cite{0709.1517,0806.2415,1007.1517}. However, the NADE and HDE
models are very successful in explaining the cosmic acceleration and
in fitting the observation data. The tension between WGC and the two
models makes us nervous. However, we find that a key assumption that
is used to motivate Eq.(\ref{deltaPhiMp}) contradicts the basic
motivation of the HDE and NADE models. Then following the analysis
of Huang \cite{0703071,0706.2215}, we find that in the HDE and NADE
models, Eq.(\ref{deltaPhiMp}) should be replace by the new
criterions (\ref{newDeltaPhiHDE}) and (\ref{newDeltaPhiNADE})
respectively.

Then we apply Eq.(\ref{newDeltaPhiNADE}) to survey the NADE model,
and find that the NADE model is consistent with the new criterion
(\ref{newDeltaPhiNADE}) within the observational constraint
\cite{0904.0928,1011.6122}. So the tension between WGC and the NADE
model does not exist if the new criterion (\ref{newDeltaPhiNADE}) is
used.

However, when applying Eq.(\ref{newDeltaPhiHDE}) to survey the HDE
models, we find that WGC cannot be satisfied in the HDE model
without interaction and spatial curvature, and it is only in the HDE
model with spatial curvature and interaction $Q=Q_3$ that
Eq.(\ref{newDeltaPhiHDE}) can be satisfied. The conclusion is
similar to what is obtained in \cite{1007.1517}, but the parameter
range within which WGC is satisfied becomes larger. Here we note
that the reason why Eq.(\ref{newDeltaPhiHDE}) can not be satisfied
in the HDE model with spatial curvature and interaction $Q=Q_1$ or
$Q=Q_2$ is that in the two types of models, only the case of $b\le0$
is regarded as the relativistic physical situation since positive
$b$ will lead $\rho_m$ to be negative in the far future in the two
models. At the same time, it can be easily checked that the negative
$b$ would make it more difficult for Eq.(\ref{HDEWGC}) to be
satisfied than the case of $b=0$. Fortunately, in the model with
$Q=Q_3$, there is no such a restriction and we can take $b>0$. This
makes it possible for us to find the combinations of
$\Omega_{m0},\Omega_{k0},b$ and $c$ satisfying WGC in the model with
$Q=Q_3$.

Summarily, in the note we suggest two new criterions to alleviate
the tension between WGC and the modes of HDE and NADE. We find that
the contradiction between WGC and the NADE model is removed when
using the new criterion (\ref{newDeltaPhiNADE}). But, in the HDE
model, we need involve the effect of the spatial curvature and chose
the interaction term as $Q=Q_3$, in order to match the new criterion
(\ref{newDeltaPhiHDE}).

\section*{Acknowledgments}
This work is supported by the Natural Science Foundation of the
Northwest University of China under Grant No. 09NW27.


\begin{thebibliography}{99}

\bibitem{kklt}
S. Kachru, R. Kallosh, A. Linde and S. P. Trivedi, Phys. Rev. D 68,
046005 (2003), [hep-th/0301240].

\bibitem{landscape}L. Susskind,
hep-th/0302219.

\bibitem{vafa}C. Vafa,
hep-th/0509212.

\bibitem{wgc1}N. Arkani-Hamed, L. Motl, A. Nicolis and C. Vafa, JHEP 0706, 060
(2007), [hep-th/0601001].

\bibitem{wgc2} H. Ooguri and C. Vafa, Nucl. Phys. B 766, 21 (2007),
[hep-th/0605264].

\bibitem{0703071}Q. G. Huang,
JHEP 0705, 096 (2007), [hep-th/0703071].

\bibitem{0706.2215}Q. G. Huang,
Phys. Rev. D 76, 061303 (2007), arXiv:0706.2215[hep-th].


\bibitem{0708.2760}Q. G. Huang,
Phys. Rev. D 77, 103518 (2008), arXiv:0708.2760 [astro-ph].

\bibitem{0710.1406}X. Wu and Z. H. Zhu, Chin. Phys. Lett. 25, 1517
(2008), arXiv:0710.1406[astro-ph].


\bibitem{0709.1517}Y. Z. Ma and X. Zhang, Phys. Lett. B 661, 239
(2008), arXiv:0709.1517[astro-ph].

\bibitem{0806.2415}X. Chen, J. Liu and Y. Gong, Chin. Phys. Lett. 25, 3086 (2008), arXiv:0806.2415[gr-qc].

\bibitem{1007.1517}C. Y. Sun,
arXiv:1007.1517[gr-qc].

\bibitem{1005.2466}X. L. Liu, J. Zhang and X. Zhang,
(2010), arXiv:1005.2466[gr-qc].

\bibitem{0811.1333}E. N. Saridakis,
Phys. Lett. B 676, 7 (2009), arXiv:0811.1333[hep-th].

\bibitem{HDE}
M. Li, Phys. Lett. B 603, 1 (2004), [hep-th/0403127].

\bibitem{HQDE}X. Zhang,
Phys. Lett. B 648, 1 (2007), [astro-ph/0604484]; J. Zhang, X. Zhang
and H. Liu,
Phys. Lett. B 651, 84 (2007), arXiv:0706.1185 [astro-ph]; X. Zhang,
Phys. Rev. D 74, 103505 (2006), [astro-ph/0609699];

\bibitem{0708.0884}H. Wei and R. G. Cai,
Phys. Lett. B 660, 113 (2008), arXiv:0708.0884[astro-ph].

\bibitem{HP}G. ¡¯t Hooft, arXiv:gr-qc/9310026; L. Susskind, J. Math. Phys. 36, 6377 (1995).

\bibitem{0403052}S. D. H. Hsu,
Phys. Lett. B 594, 13 (2004), arXiv:hep-th/0403052.

\bibitem{0904.0928}M. Li, X. D. Li, S. Wang and X. Zhang,
JCAP 0906, 036 (2009), arXiv:0904.0928[astroph.CO].

\bibitem{1011.6122}Y. H. Li, J. Z. Ma, J. L. Cui, Z. Wang and X.
Zhang,
arXiv:1011.6122[astro-ph.CO]


\bibitem{0910.3855}
M. Li, X. D. Li, S. Wang, Y. Wang and X. Zhang,
JCAP 0912, 014 (2009), arXiv:0910.3855[astro-ph.CO].


\bibitem{0404229}Q. G. Huang and M. Li,
JCAP 0408, 013 (2004), [arXiv:astro-ph/0404229].

\end{thebibliography}
\end{document}